\documentclass[english]{article}
\usepackage[utf8]{inputenc}
\usepackage[T1]{fontenc}

\usepackage{listings}
\usepackage{babel}
\usepackage{amsmath}
\usepackage{graphicx}
\usepackage{fancyhdr}
\usepackage{color}
\usepackage{caption}
\usepackage{tabularx}
\usepackage{array,tabularx,booktabs}
\usepackage{multirow}
\usepackage[table, dvipsnames]{xcolor}
\usepackage{caption}
\usepackage{subcaption}
\usepackage{authblk}

\colorlet{punct}{red!60!black}
\definecolor{background}{HTML}{EEEEEE}
\definecolor{delim}{RGB}{20,105,176}
\colorlet{numb}{black}

\lstdefinelanguage{json}{
    basicstyle=\normalfont\ttfamily,
    numbers=left,
    numberstyle=\scriptsize,
    stepnumber=1,
    numbersep=8pt,
    showstringspaces=false,
    breaklines=true,
    frame=lines,
    backgroundcolor=\color{background},
    literate=
     *{0}{{{\color{numb}0}}}{1}
      {1}{{{\color{numb}1}}}{1}
      {2}{{{\color{numb}2}}}{1}
      {3}{{{\color{numb}3}}}{1}
      {4}{{{\color{numb}4}}}{1}
      {5}{{{\color{numb}5}}}{1}
      {6}{{{\color{numb}6}}}{1}
      {7}{{{\color{numb}7}}}{1}
      {8}{{{\color{numb}8}}}{1}
      {9}{{{\color{numb}9}}}{1}
      {:}{{{\color{punct}{:}}}}{1}
      {,}{{{\color{punct}{,}}}}{1}
      {\{}{{{\color{delim}{\{}}}}{1}
      {\}}{{{\color{delim}{\}}}}}{1}
      {[}{{{\color{delim}{[}}}}{1}
      {]}{{{\color{delim}{]}}}}{1},
}

\newlength{\subcolumnwidth}

\newcommand{\nextsubcolumn}[1][]{%
  \cr\noalign{\hfill}
  \if\relax\detokenize{#1}\relax\else\hsize=#1\setlength{\subcolumnwidth}{\hsize}\fi
}

\begin{document}

\title{An individual-level ground truth dataset for home location detection}

\author{Luca Pappalardo\textsuperscript{1}, Leo Ferres\textsuperscript{2, 4}, Manuel Sacasa\textsuperscript{2, 4},\\ Ciro Cattuto\textsuperscript{3}, Loreto Bravo\textsuperscript{2, 4}
}

\maketitle

\thispagestyle{fancy}
1. ISTI-CNR, Pisa, Italy; 2. Faculty of Engineering, Universidad del Desarrollo, Santiago, Chile; 3. University of Turin and ISI Foundation, Torino, Italy; 4. Telef\'onica R\&D Santiago, Chile



\begin{abstract}
Home detection, assigning a phone device to its home antenna, is a ubiquitous part of most studies in the literature on mobile phone data. Despite its widespread use, home detection relies on a few assumptions that are difficult to check without ground truth, i.e., where the individual that owns the device resides.
In this paper, we provide an unprecedented evaluation of the accuracy of home detection algorithms on a group of sixty-five participants for whom we know their exact home address and the antennas that might serve them.
Besides, we analyze not only Call Detail Records (CDRs) but also two other mobile phone streams: eXtended Detail Records (XDRs, the ``data'' channel) and Control Plane Records (CPRs, the network stream).
These data streams vary not only in their temporal granularity but also they differ in the data generation mechanism', e.g., CDRs are purely human-triggered while CPR is purely machine-triggered events.
Finally, we quantify the amount of data that is needed for each stream to carry out successful home detection for each stream.
We find that the choice of stream and the algorithm heavily influences home detection, with an hour-of-day algorithm for the XDRs performing the best, and with CPRs performing best for the amount of data needed to perform home detection. Our work is useful for researchers and practitioners in order to minimize data requests and to maximize the accuracy of home antenna location.

\end{abstract}

\section{Background \& Summary}
\label{sec:introduction}

Nowadays, there is a strong demand by governments, public institutions, and statistics bureaus for inferring sociodemographic information from digital data in support of official statistics, to reduce costs and time of data collection and increase the timeliness and scalability of the collected data \cite{struijs2014official, un1, un2}.
In particular, the recent years have witnessed the emergence of methodologies that use heterogeneous digital data - such as social media data, GPS traces, or mobile phone records - to estimate socio-economic indicators of objective and subjective well-being \cite{voukelatou2020measuring,  gao2019computational, piaggesi2019predicting, luo2017inferring, pappalardo2016analytical,  llorente2015social, vscepanovic2015mobile, pappalardo2015using, marchetti2015small, mao2015quantifying, frias2012relation, eagle2010network}.
Among the many applications of data science to official statistics, the usage of mobile phone data to identify \emph{where someone lives}, i.e., detecting an individual's home location, has emerged as one of the most critical issues in many tasks, such as the estimation of population density \cite{deville2014dynamic, khodabandelou2016population, salat2020method}, commuting and migration flows \cite{gabrielli2015city, lai2019exploring, hankaew2019inferring, blumenstock2012inferring}, air pollution \cite{picornell2018population, li2019dynamic}, and the estimation of privacy risk \cite{demontjoye2013unique, pellungrini2017data, demontoye2018privacy, pellungrini2020modeling, fiore2019privacy}.
The knowledge of individuals' home location forms the crucial link between digital data and census data, making it a key enabler for the integration of these two sources of information.

Most of the home detection algorithms (HDAs) proposed in the literature \cite{ahas2010using, frias-martinez2010towards, vanhoof2018assessing, vanhoof2020performance} process mobile phone records according to heuristics rather than principled approaches.
Indeed, they rely on simple decision rules based on how much, and when, an individual calls in each location during a period of observation.

The simplest HDA identifies an individual's home location as the one in which they made the highest number of calls.
A variant of this HDA identifies an individual's home location as the one in which they make the highest number of calls during nighttime (between 7pm and 9am).
Other HDAs use a combination of criteria or slight variations of the ones mentioned above \cite{vanhoof2018assessing, vanhoof2020performance}.
Although these HDAs have been used in many works and tools \cite{phithakkitnukoon2012socio, pappalardo2016analytical, pappalardo2015returners, cho2011friendship, cuttone2018understanding, wang2011human, eagle2010network, pappalardo2020scikitmobility}, a thorough validation of their accuracy still lacks in the literature.

On the one hand, HDAs have been validated on human-triggered data such as Call Detail Records (CDRs), which describe each user's position only when they make or receive a call.
Since the inter-event time between two calls is bursty \cite{song2010modelling}, CDRs are sparse and provide an incomplete picture of an individual's positions over time \cite{barbosa2018human}.
It is not clear whether human/machine-triggered records (e.g., eXtended Detail Records or XDRs) or pure network machine-triggered records (e.g., Control Planes Records or CPRs) can overcome the limitations of using the more temporally sparse CDRs and provide more accurate estimates of an individual's home location.

On the other hand, there is no consensus on which HDA is the most accurate and what determines its accuracy, mainly because of the lack of proper validation.
Apart from sporadic cases \cite{ahas2010using, frias-martinez2010towards}, ground truth data at the individual level are not provided by mobile providers for privacy protection reasons, making it difficult to obtain a large sample of users for which complete information about positions and residence are available at the same time.
For example, Vanhoof et al. \cite{vanhoof2018assessing} provide a high-level validation of the most popular HDAs by comparing each mobile phone tower’s population as estimated by official censuses with the number of users whose home is detected to be in that tower.
They conclude that there is an urgent need for validation of HDAs at the individual level, i.e., evaluating the performance in detecting the home location on a set of individuals for which the actual home location is known \cite{vanhoof2018assessing}.

This paper provides an unprecedented dataset that allows validation of HDAs on individual-level ground truth data and three streams of mobile phone records -- CDRs, XDRs, and CPRs.
Specifically, 65 users working for Telef\'onica Chile gave their written consent to provide us access to their phone records for two weeks, as well as their actual address of residence.
This information allows us to assess the accuracy of HDAs, i.e., their capacity to detect a user's actual home correctly, on a ground truth dataset.
Our validation reveals the most accurate HDA among the popular ones proposed in the literature, and that human-machine- and machine-triggered records (XDRs and CPRs, respectively) improve the accuracy of HDAs considerably with respect to pure human-triggered records (CDRs).
Moreover, we set up a ``data minimization'' experiment to study how the accuracy of detecting home locations changes by the stream used and the number of records for each user.
We find that, depending on the stream, just a small fraction of the records is enough to achieve reasonably accurate estimations of an individual's home location, hence providing a tool to manage the uncertainty and utility trade-off in geo-privacy.

With this paper, we also release both the code to reproduce the HDAs and a dataset that describes, for each ground truth user, the towers visited, the visitation frequency to each tower, and an estimation of the tower(s) associated with the user's address of residence.
To the best of our knowledge, this is an unprecedented contribution to the field, which paves the road towards the definition of more accurate home detection algorithms.

\section{Methods}
\label{sec:mobile-phone-dimens}

\begin{figure}
\centering
\begin{subfigure}{\textwidth}
\includegraphics[width=\textwidth]{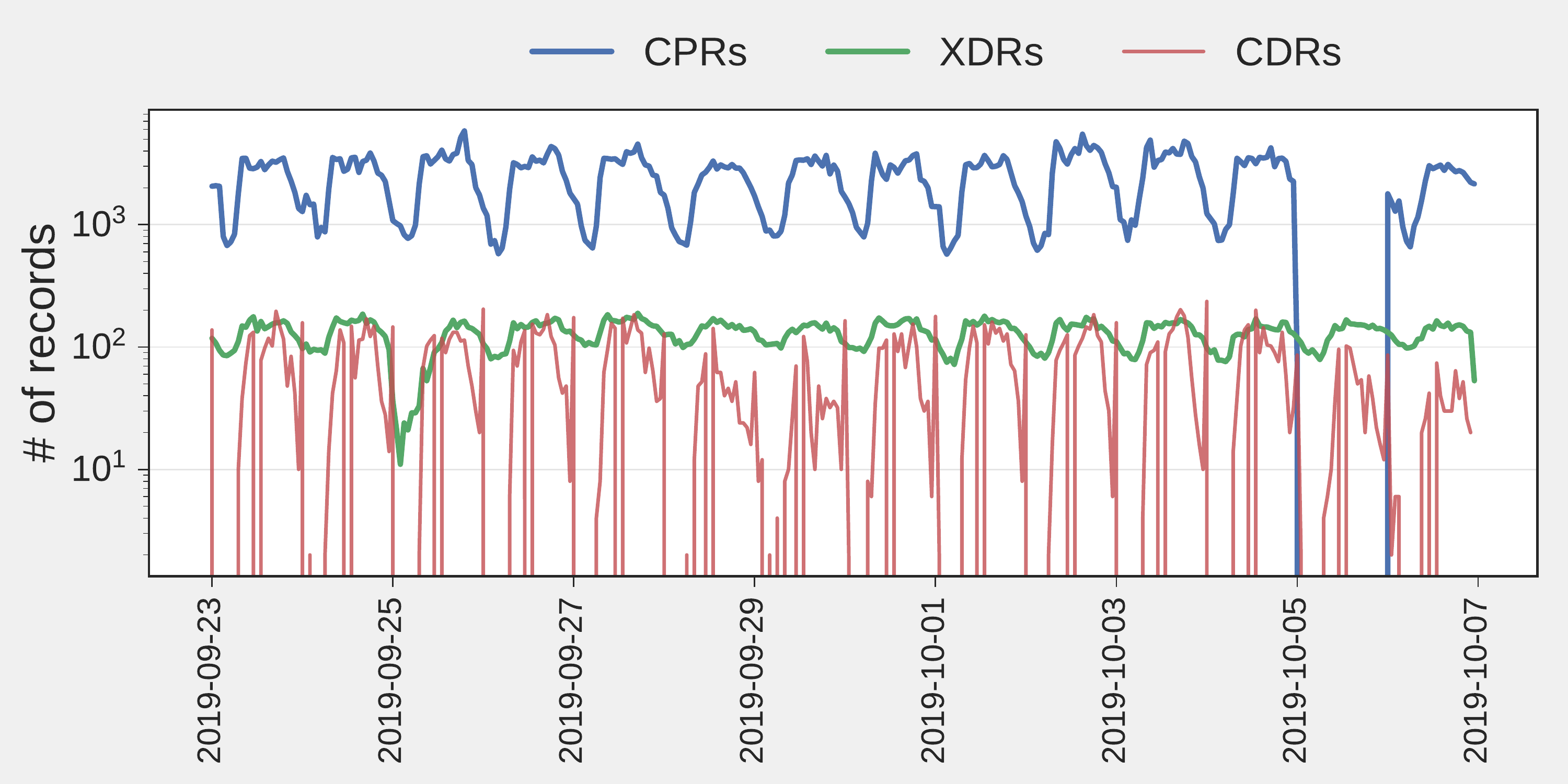}
\caption{$\log$ number of events per hour}
\end{subfigure}
\hfill
\begin{subfigure}{\textwidth}
\includegraphics[width=\textwidth]{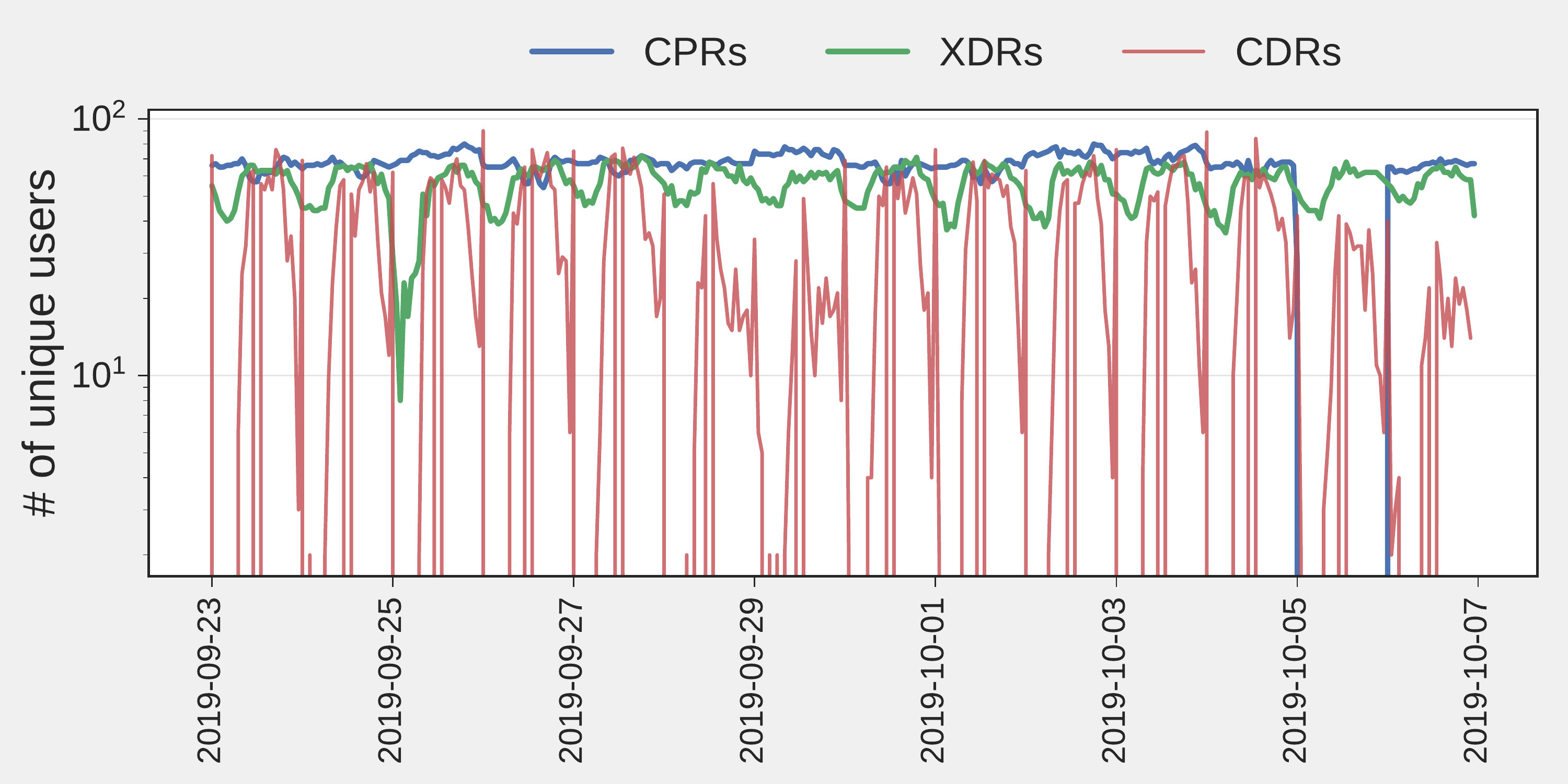}
\caption{$\log$ number of active users per hour}
\end{subfigure}
\caption{A comparison of the number of \textbf{(a)} records and \textbf{(b)} active users for CDRs, XDRs and CPRs, for 14 days (except Oct 5 for CPRs, for which there was no data ingestion).}
\label{fig:example}
\end{figure}

We obtained written consent from 65 people working for Telefónica Chile to use the precise latitude and longitude of their homes in Santiago de Chile, calculated over the original addresses (e.g., 123 Santiago Street, Providencia) with reverse lookup using Google Maps.
We also obtained consent to gather two weeks (14 days, September 24-October 6, 2019 inclusive) of their historical mobile phone records for three streams: Call Detail Records (CDRs), eXtended Detail Records (XDRs), and Control Plane Records (CPRs).
For CPRs, October 5 is missing (see Figure \ref{fig:example}), and there are 13 days only.

\begin{figure}
    \centering
    \includegraphics[width=1\linewidth]{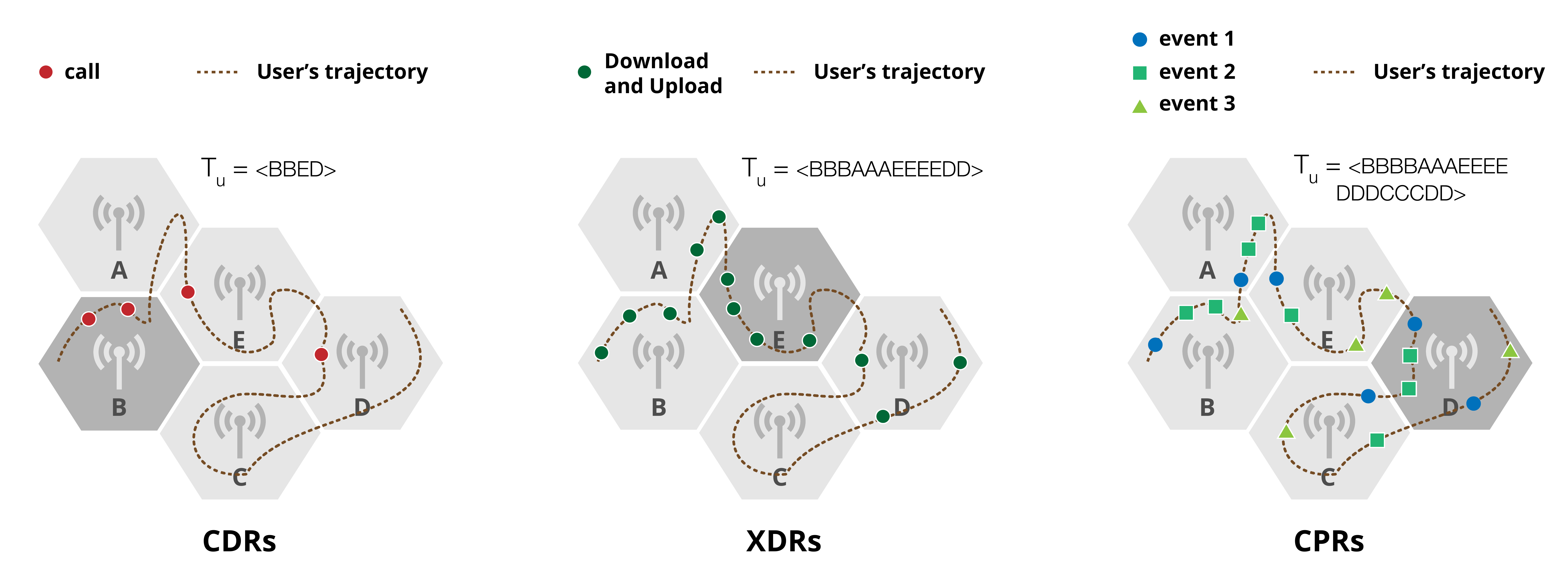}
    \caption{Illustrative example of Call Detail Records (CDRs), eXtended Detail Records (XDRs) and Control Plane Records (CPRs) of a user $u$.
    The hexagons represent mobile phone towers and black dots the positions where the user starts a call (CDRs), a download/upload operation (XDRs), or a network event (CPRs).
    Red dots indicate calls, green dots indicate download/upload operations, blue dots, green squares and green triangles indicate network events. The dotted line indicates the real movement of the user, from the left to the right.
    The dark grey tower indicates the tower in which the user has the highest number of records.
    From this example, we observe that: 1) CDRs are the most sparse and CPRs the most dense; 2) when $u$ do not perform any call or download/upload operation in the area covered by a tower, there is no information about the user's position in CDRs and XDRs; 3) CPRs is the only type of records that can describe all the towers $u$ passed through.
    }
    \label{fig:example_CDRs_XDRs_CPRs}
\end{figure}

Mobile phone operators collect CDRs and XDRs for billing and operational purposes.
CDRs are purely user-triggered, i.e., they are generated by the users every time they make or receive a phone call.
Consequently, when a user does not make/receive a call, their position is not recorded.
XDRs are a mixture of human- and device-triggered, either by explicitly requesting an {\tt http} address or automatically downloading content from the Internet (e.g., emails).
In contrast, CPRs are network-triggered (e.g., assigning a new antenna, connecting new devices) and are used to monitor the cellphone network health.
Although all three streams are equal in terms of their geo-location properties (the set of towers remains more or less constant in time), they vary significantly in their time granularity and data sparsity.
Figure \ref{fig:example_CDRs_XDRs_CPRs} shows an illustrative example of CDRs, XDRs, and CPRs of an individual, and how the most frequented tower changes stream by stream.

\begin{figure}
\centering
\includegraphics[width=8cm]{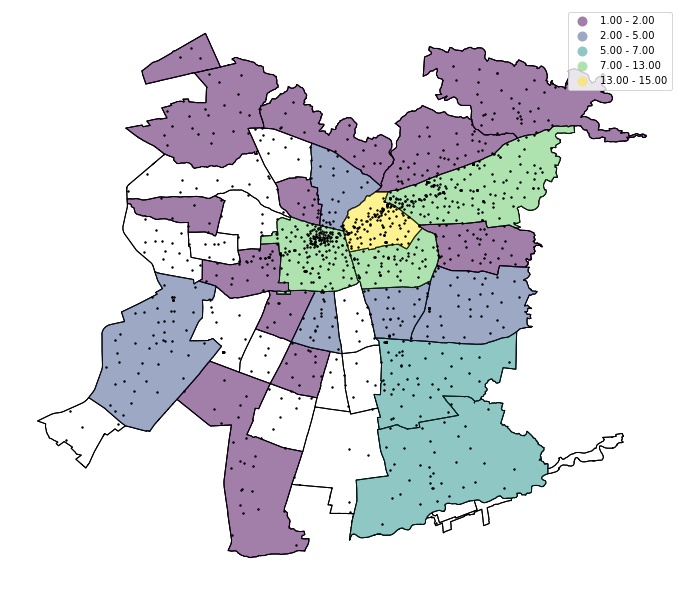}
\caption{Antennas (black dots) and general spatial location of our 100 subscribers in the largest urban area of Santiago de Chile. Antennas cluster in regions with more demand, and become sparser in regions with less. The noticeable line of antennas in the middle of the map are those that fall in the wealthier/financial comunas (a large administrative unit, represented in our map with black lines).}
\label{fig:antennas}
\end{figure}

Formally, a Call Detail Record (CDR) is a tuple $(u_o,u_i,t,d,A_o,A_i)$, where $u_o$ and $u_i$ are the identifier of the user that places the call (the caller) and the one that receives the call (the callee), respectively; $t$ is a timestamp of when the call was placed, and $d$ is the call duration (in minutes).
$A_o$ and $A_i$ are the antennas for the ``outgoing call'' (where the call is placed) and the ``receiving call'' (where the call is picked up).
Our dataset contains 1933 antennas (Figure \ref{fig:antennas}), which may be a tower (containing many antennas), or a single antenna, as in indoor spaces \cite{Beiro2018}.
In this study, an antenna $A$ is associated with its location in space expressed as a latitude and longitude pair, hence $A=(l_x, l_y)$ where $l_x$, and $l_y$ are the longitude and latitude of $A$.
For simplicity, from now on, we always use the term ``tower'' when referring to a location a user is located at.

An eXtended Detail Record (XDR) is a tuple $(u,t,A,k)$, in which there is only one antenna $A$ involved, $u$ is the caller's identifier, $t$ is a timestamp of when the record is created, and $k$ is the amount of downloaded information (in kilobytes).
A Control Plane Record (CPR) is a tuple $(u,t,A, e)$, where each $e \in E$ is an ``event'' of the network.
There are many possible control-plane events, such as ``handovers'' (when a new antenna is pushed on the device).

For the period under investigation, we have 19,234 CDRs, 43,607 XDRs and 772,871 CPRs.
Figure \ref{fig:example}a compares the number of records in our dataset of the three streams.
CPRs (network-triggered) are by far the most frequent ones (i.e., most records per user), followed by XDRs (human- and device-triggered) and CDRs (human-triggered).
CDRs are also the data type with the highest variance across the days of the week.
Figure \ref{fig:example}b shows the number of active users (those with at least one record) per hour in our dataset for each stream.
Again, CPRs show the highest number of active users, followed by XDRs and CDRs.
Note also that CPRs have the lowest variance between the week's days, while CDRs have the highest variance.

\subsection*{Code availability}
The code to reproduce the analyses in the paper is available at\\ \texttt{https://github.com/leoferres/home\_loc}.


\section{Data Records}

From the CDRs, XDRs, and CPRs of the users, we extract aggregated data about the activity in each tower according to five home detection algorithms (HDAs, see Section \ref{sec:validation}).
Tables \ref{tab:activity_data}, \ref{tab:tower_data} and \ref{tab:ground_truth} show the structure of the released data.
In practice, the tower identified by an HDA $X$ on a stream $T$ as the home location of a user $u$ is the tower in the row with the highest value of ``activity'' in Table \ref{tab:activity_data} in which ``data type''$=T$ and ``HDA''$=X$.
We indicate as $H_{X, T}(u)$ the tower detected as the home location by HDA $X$ on stream $T$ for user $u$.

\subsection*{Activity}
In the \texttt{Activity} dataset, each row describes the activity a user has in a tower, according to a data stream and an HDA.
Column ``device'' (type: string) indicates the (anonymized) device identifier.
Column ``tower'' (type: string) indicates the identifier of a mobile phone tower. Column ``activity'' (type: integer) indicates the amount of activity the user has in the tower. Column ``stream'' (type: string) indicates the type of records (CDRs, XDRs, or CPRs) and column ``HDA'' the home detection algorithm.
Given a stream $T$, a user $u$, and an HDA $X$, the row with the highest value of ``activity'' is the home detected by $X$ on $T$ for $u$.
For example, for CDRs and HDA1, column ``activity'' indicates the number of CDRs the user has for that tower.
For XDRs and HDA2, the same column indicates the number of distinct days the user has at least one XDR for that tower.
Table \ref{tab:activity_data} shows some examples of records in the \texttt{Activity} dataset corresponding to a single user.
Note that the records are sorted by device, stream, HDA, and activity.
In total, the \texttt{Activity} dataset contains 260,400 rows.

\begin{table}[htb!]\centering
\newcolumntype{Y}{>{\centering\arraybackslash}X}
    \begin{tabularx}{\linewidth}{YYYYYYYYYY}
    \toprule

    \textbf{device} & \textbf{tower} & \textbf{activity} & \textbf{stream} & \textbf{HDA} \\
    \midrule
afa64           & ESALT          & 5                 & CDRs               & HDA1        \\
afa64           & \_0056         & 3                 & CDRs                & HDA1        \\
afa64           & SALAL          & 1                 & CDRs                & HDA1        \\
afa64           & ESALT          & 2                 & CDRs                & HDA2        \\
afa64           & \_0056         & 1                 & CDRs                & HDA2       \\
\dots & \dots & \dots & \dots & \dots \\
    \bottomrule
    \end{tabularx}

    \caption{Structure of the activity dataset.
    }
    \label{tab:activity_data}
\end{table}

\subsection*{Towers}
In the \texttt{Towers} dataset, each row describes the identifier and the geographic location of a mobile phone tower.
Column ``tower'' (type: string) indicates the identifier of a mobile phone tower. Columns ``lat'' (type: float) and ``lng'' (type: float) indicate the tower's latitude and longitude, respectively.
Table \ref{tab:tower_data} shows some examples or records describing mobile phone towers.
In total, the \texttt{Towers} dataset containts 65 rows.

\begin{table}[htb!]\centering
\newcolumntype{Y}{>{\centering\arraybackslash}X}
    \begin{tabularx}{\linewidth}{YYY}
    \toprule

    \textbf{tower} & \textbf{lat} & \textbf{lng} \\
    \midrule
ESALT          & -33.40374    & -70.63715    \\
LUISZ          & -33.57250    & -70.57569    \\
SUEG1          & -33.48468    & -70.55035    \\
AGSTF          & -33.48468    & -70.55035    \\
PAROC          & -33.44548    & -70.61918 \\
\dots & \dots & \dots \\
    \bottomrule
    \end{tabularx}

    \caption{Structure of the towers dataset.
    }
    \label{tab:tower_data}
\end{table}

\subsection*{Ground truth}
In the \texttt{Ground truth} dataset, each row describes a mobile phone device and the identifiers of the three closest tower to the user's actual home location.
Column ``device'' (type: string) indicates the (anonymized) device identifier.
Columns ``closest'', ``2nd closest'', and ``3rd closest'' (type: string) indicate the identifiers of the towers that are the closest tower, the second closest tower, and the third closest tower to the actual home location of the user, respectively.
Table \ref{tab:ground_truth} shows some examples of ground truth records.
In total, we have 65 rows in the \texttt{Ground truth} dataset.

\begin{table}[htb!]\centering
\newcolumntype{Y}{>{\centering\arraybackslash}X}
    \begin{tabularx}{\linewidth}{YYYY}
    \toprule

    \textbf{device} & \textbf{closest} & \textbf{2nd closest} & \textbf{3rd closest} \\
    \midrule
afa64           & ANTPR            & MEINS                & RECC1                \\
214ab           & ELTRO            & JUMPE                & LPRES                \\
1c75db          & DIGOR            & DGAC1                & SANT2                \\
f1599           & LOSEN            & AMCLF                & PIRF2                \\
d666d           & ANTPR            & APERR                & MEINS   \\
\dots & \dots & \dots & \dots \\
    \bottomrule
    \end{tabularx}

    \caption{Structure of the ground truth dataset.
   }
    \label{tab:ground_truth}
\end{table}

\section{Technical Validation}
\label{sec:validation}

We use the released dataset to assess the agreement between popular home detection algorithms and assess their accuracy with respect to the ground truth dataset.

We consider the five most popular home detection algorithms defined for mobile phone data \cite{vanhoof2018assessing}, denoted as HDA1-HDA5.
For every stream - CDRs, XDRs, and CPRs - and for every user $u$, the HDAs calculate the most active tower given some constraints.
Algorithm HDA1 counts the number of records at every tower for $u$.
The tower with the highest activity is $u$'s home location.
Algorithm HDA2 calculates the activity of towers for every \textit{unique} day. In our case, for CDRs and XDRs, if the user spends every day at home, the highest number will be 14, while for CPRs, it will be 13.
Thus, wherever $u$ is on most days over two weeks is their home.
Algorithm HDA3 calculates the most active towers during nighttime: between 7pm and 7am, inclusive.
The tower that is most active during this period at night is $u$'s home location.
Algorithm HDA4 is an extension of HDA1 that implements a spatial perimeter of $1$~km around each tower.
Whatever tower aggregates most activities in the perimeter is $u$'s home.
Finally, according to HDA5, $u$'s home location is the tower with most activities in a $1$~km perimeter between 7pm and 7am.\footnote{The code of the HDAs can be found at the following GitHub repository: {\tt https://github.com/leoferres/home\_loc}.}

\subsection{Agreement between HDAs}
We investigate whether, for the same user, different HDAs detect the same home location and how the choice of the stream influences the agreement between algorithms.
Given a stream $T \in \{\mbox{CDRs}, \mbox{XDRs}, \mbox{CPRs}\}$, we assess to which degree two HDAs $X$ and $Y$ agree on a set of individuals by evaluating the Simple Matching Coefficient (SMC) \cite{vanhoof2018assessing}, defined as:
\begin{equation}
\mbox{SMC}_T(X, Y) = 100 * \frac{\sum_{i=u}^N \delta(H_{X, T}(u), H_{Y, T}(u))}{N}
\end{equation}
where $u=1 \dots N$ denotes the $N$ users in our dataset, $H_{X, T}(u)$ and $H_{Y, T}(u)$ denote the home location detected for user $u$ on stream $T$ by HDAs $X$ and $Y$ respectively, and $\delta$ is the Kronecker delta which is equal to 1 when $H_{X, T}(u) = H_{Y, T}(u)$ and 0 otherwise.
Values of SMC thus range between 0 and 100.
They can be interpreted as the percentage of individual cases for which both algorithms detected the same home location, i.e., the agreement between two HDAs.

\begin{figure}
    \centering
    \begin{subfigure}[b]{\textwidth} 
                \includegraphics[width=0.45\textwidth]{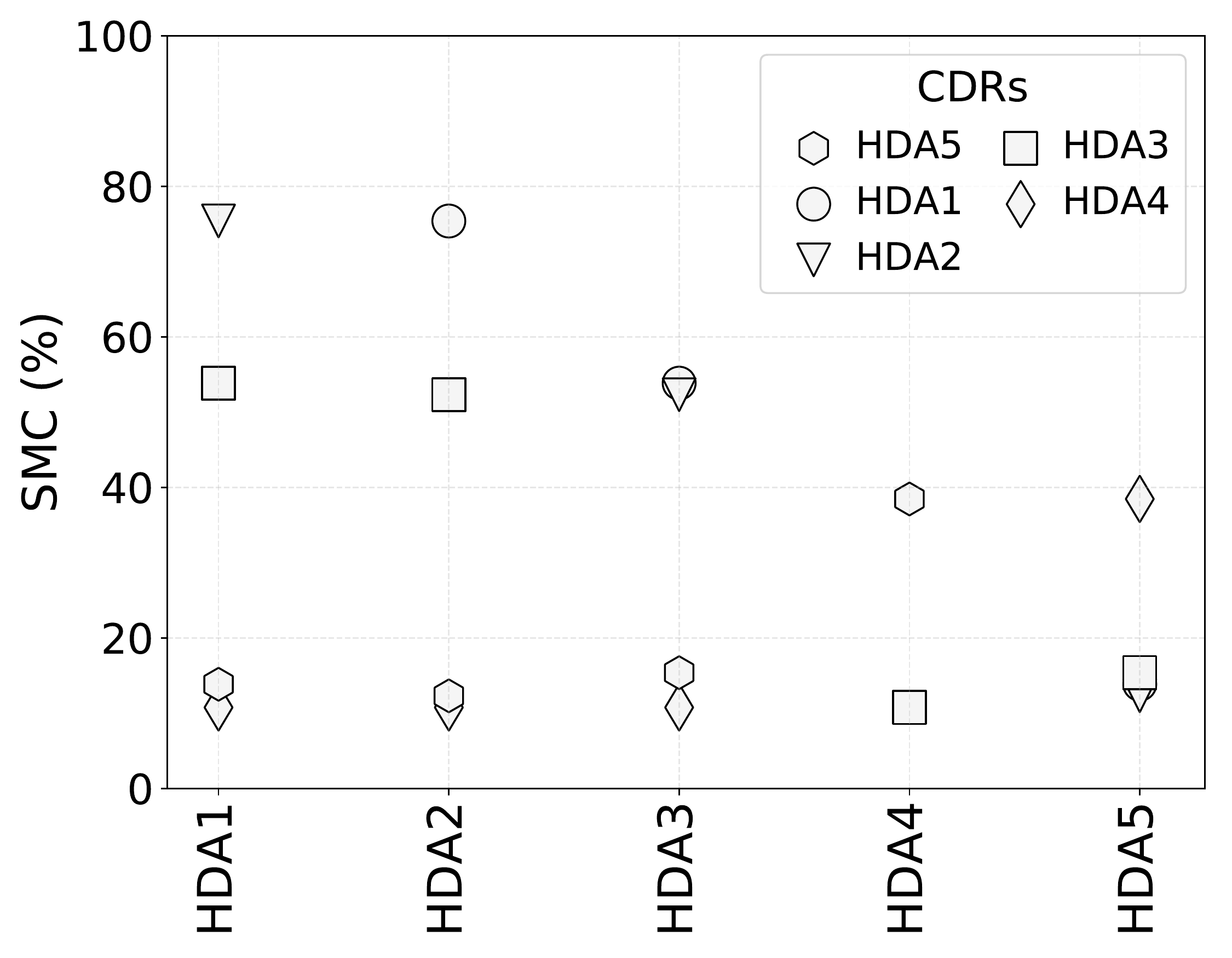}
                \includegraphics[width=0.45\textwidth]{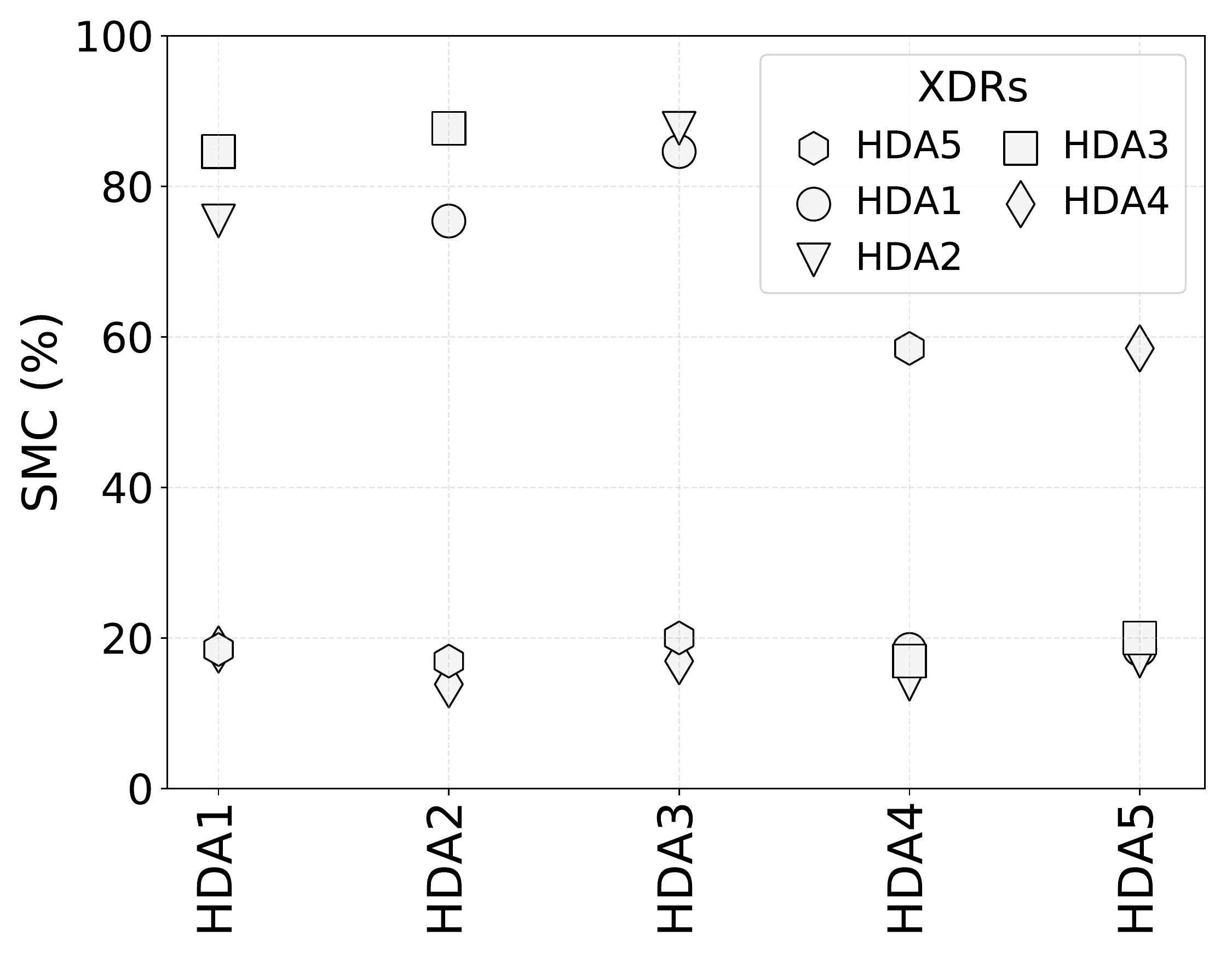}
                \includegraphics[width=0.45\textwidth]{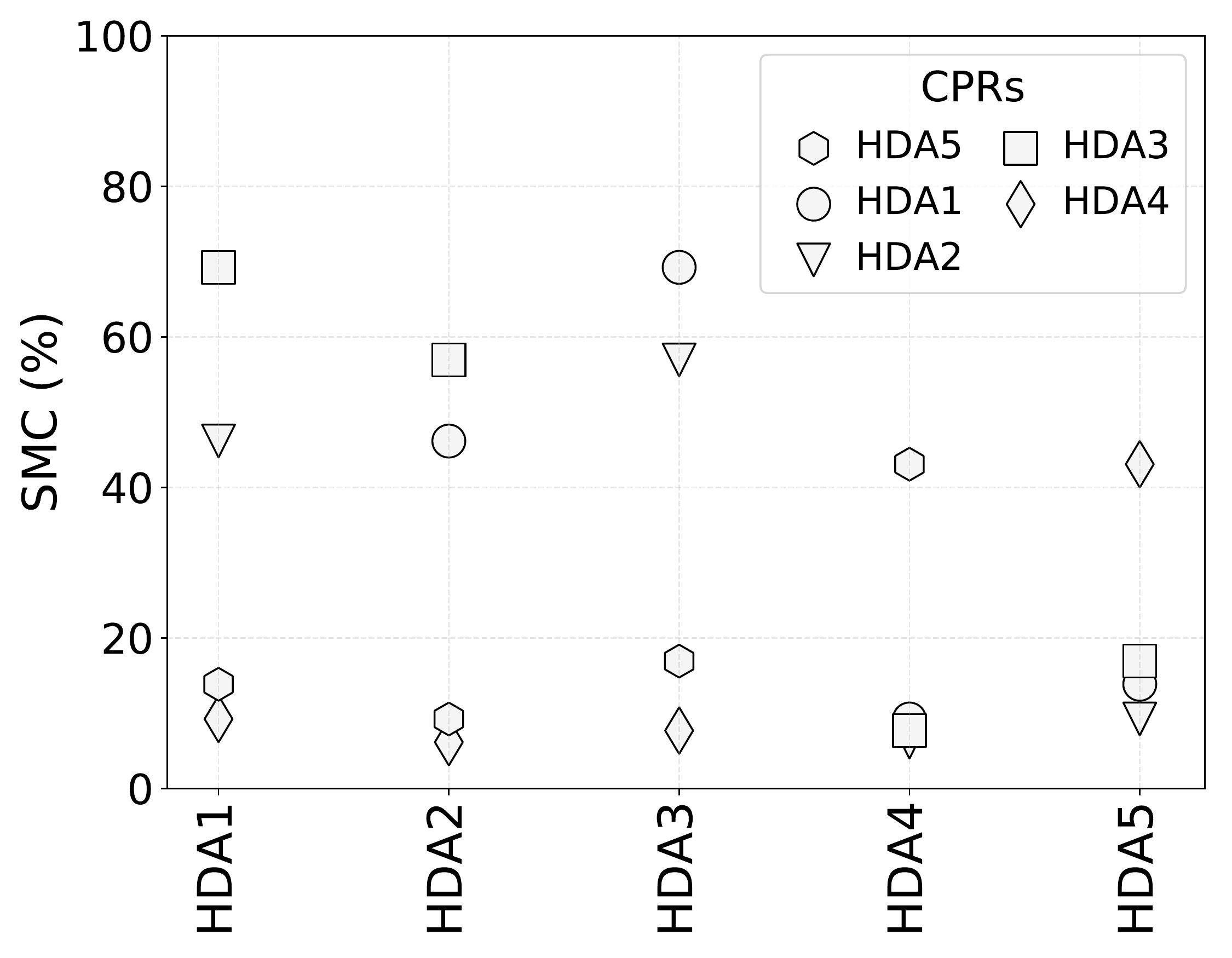}
                \hfill
                \includegraphics[width=0.45\textwidth]{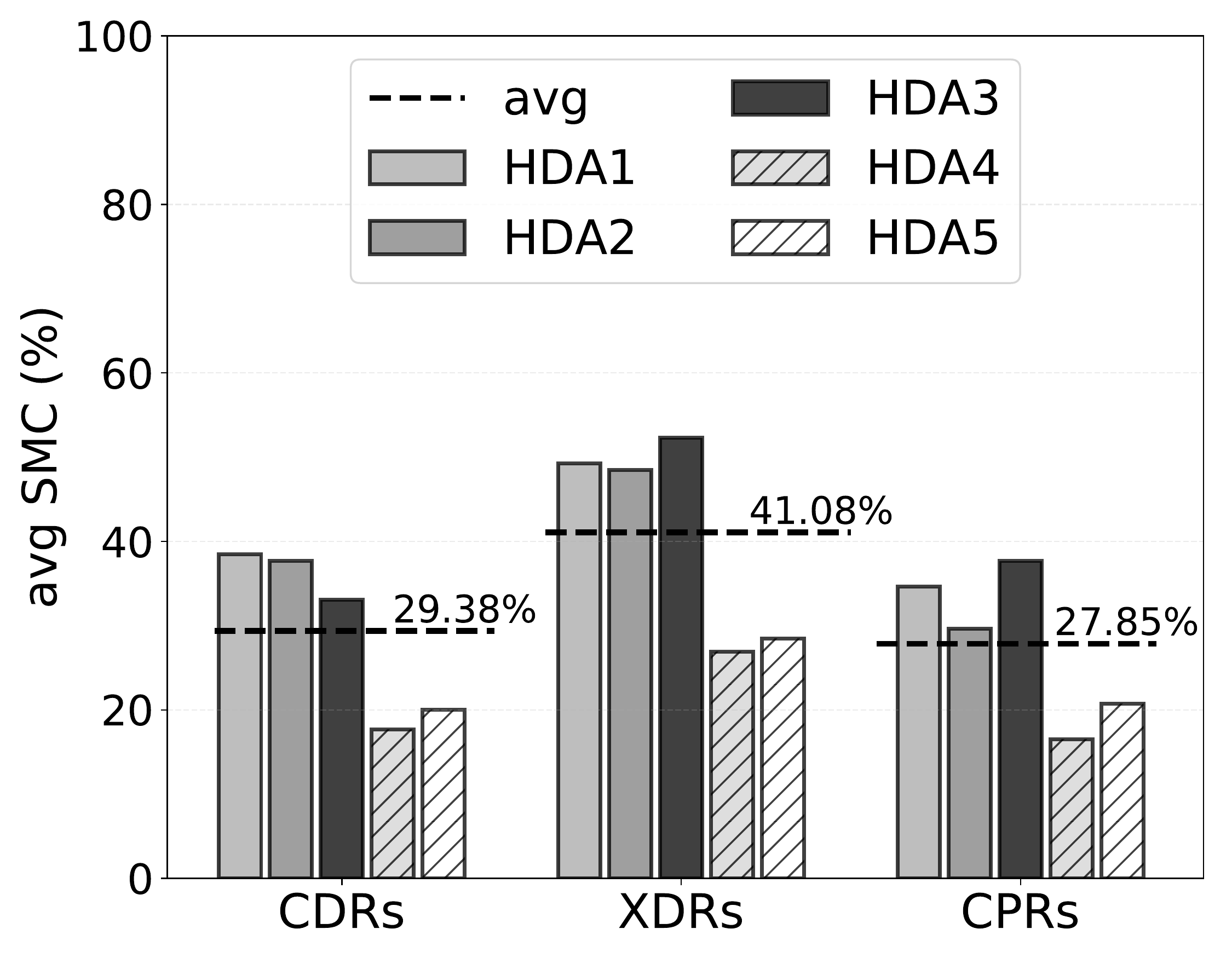}
        \end{subfigure}
    \caption{The $SMC$ of each HDA with the other HDAs, for CDRs (upper left), XDRs (upper right) and CPRs (bottom left). XDRs have the highest $SMC$ values, CPRs the second highest $SMC$ values. (Bottom right) Average SMC of each pair of HDAs, for each data type. The red dashed line indicates the average SMC of a data type.}
    \label{fig:smc_hdas}
\end{figure}

Figure \ref{fig:smc_hdas} shows the SMC for each combination of HDAs and stream and the average SMC between each HDA and all the others, for each stream.
We find that the average agreement between the HDAs is the highest for XDRs (41.09\%) and the lowest for CPRs (27.85\%).
Moreover, there are two distinct groups of HDAs with similar SMC: HDA1, HDA2, and HDA3 on one side, and HDA4 and HDA5 on the other side (Figure \ref{fig:smc_hdas}), regardless of the stream.

Our results provide important practical insights.
First, XDRs should be preferred when performing home location detection.
Since they lead to the highest agreement between HDAs, XDRs would guarantee a higher reproducibility of the results.
Unfortunately, most of the works in the literature use CDRs, and with different HDAs.
Second, for XDRs, the average agreement is about 41\%.
This means that in more than half of the cases, the HDAs disagree on what the user's home location should be, highlighting \emph{a fortiori} that the choice of the HDA is crucial.
The best solution would be to select the combination of HDA and stream that, on average, gives the most accurate home location detection.
Therefore, what is the most accurate combination of HDA and stream on individual ground truth data?


\subsection{Accuracy of HDAs}
For each user $u \in G$, where $G$ is the set of the 65 users, we know the address of residence, from which we obtain the exact position (e.g., latitude and longitude) of their actual home location $H^{(u)}$ using Google Maps.
Since the closest tower is not necessarily the one that serves a user at home\footnote{
In some areas of the city where the density of towers is high, e.g., close to downtown, some antennas are turned off by the operator at different times of the day, or they become so heavily used that the network re-routes some users.
Moreover, a user may have a tower close to their home, but with the antenna's azimuth pointing the opposite direction.
In contrast, one tower farther away that has an antenna directly "illuminating" the user's home.}, for each $u \in G$, we compute the three closest mobile phone towers, $H_1^{(u)}$, $H_2^{(u)}$ and $H_3^{(u)}$, to the ground truth position $H^{(u)}$.

Given an HDA $X$, a stream $T$, and a user $u$, we say that $H_{X, T}(u)$ is correct if $H_{X, T}(u) \in \{H_1^{(u)}, H_2^{(u)}, H_3^{(u)}\}$, i.e., if the home location detected by $X$ on $T$ is at least on of the three closest towers to $u$'s ground truth home location.
Therefore, we define the \emph{home detection accuracy} ACC$_{X, T}$ of the combination of HDA $X$ and stream $T$ as the number of correctly classified home locations over the total number of ground truth users $|G|$.
As an instance, ACC$_{X, T} = 0.50$ means that, using stream $T$, HDA $X$ can correctly detect the home location for half of ground truth users.

\begin{figure}
    \centering
    \begin{subfigure}{\textwidth} 
                \includegraphics[width=0.45\textwidth]{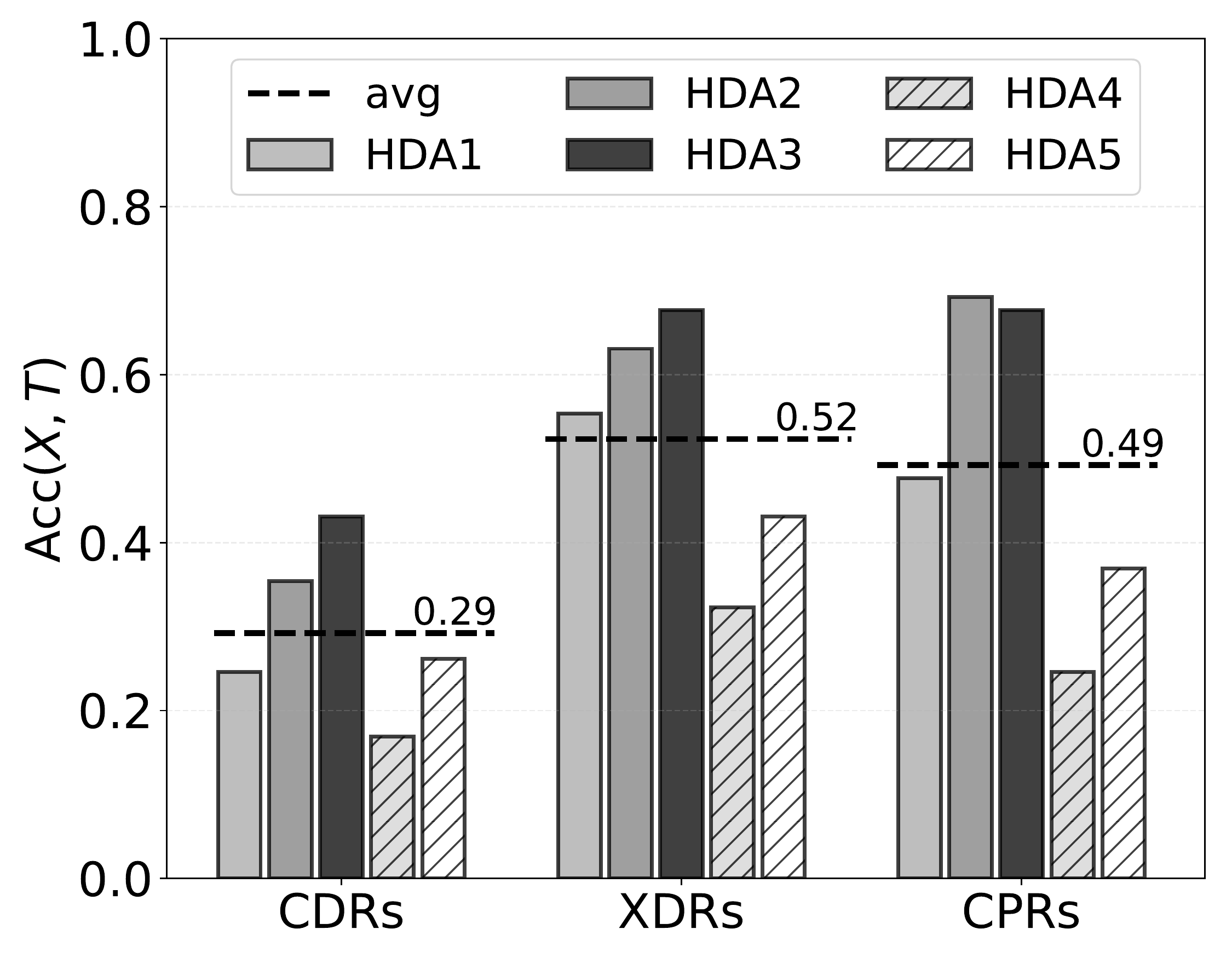}
                \includegraphics[width=0.45\textwidth]{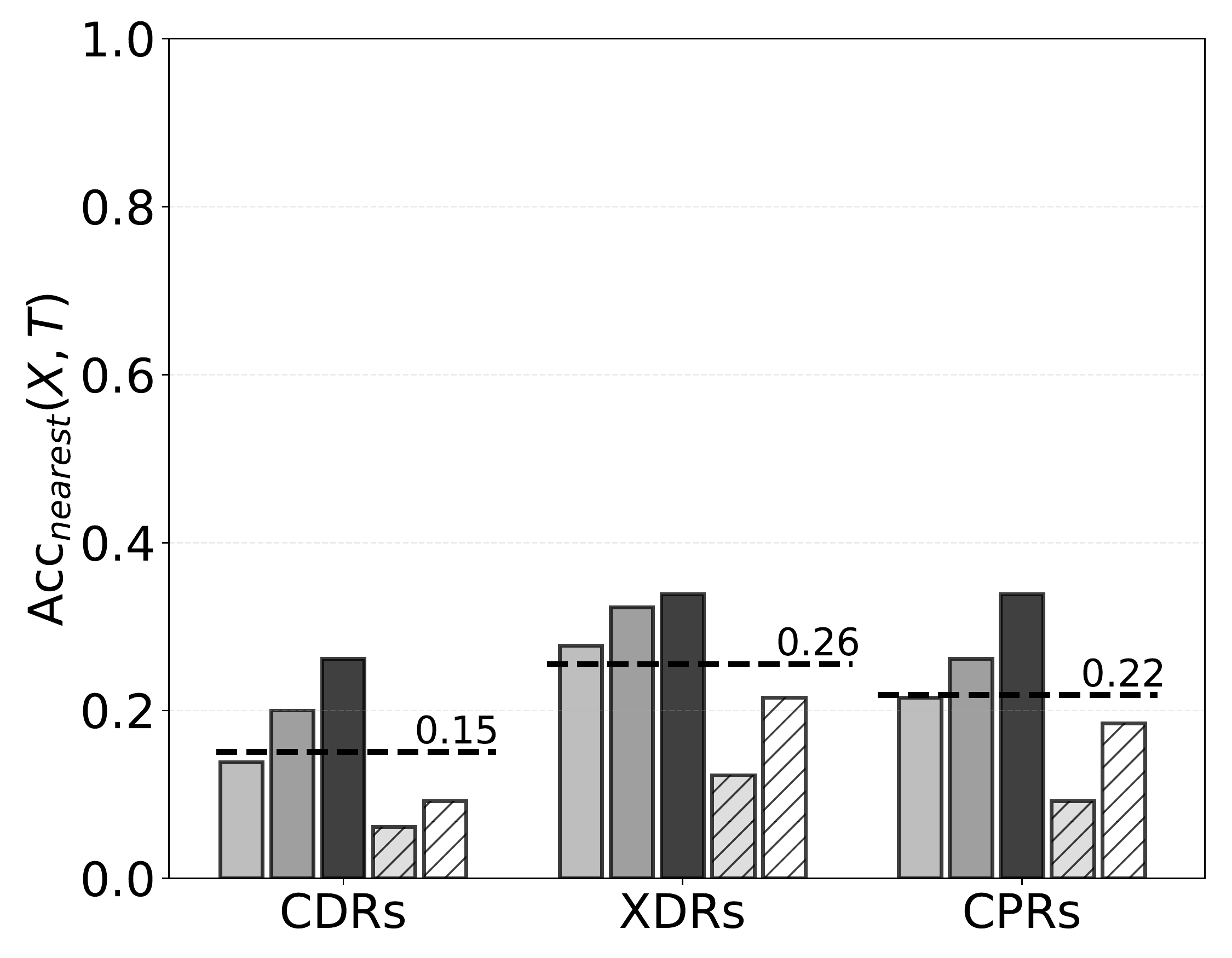}
                \includegraphics[width=0.45\textwidth]{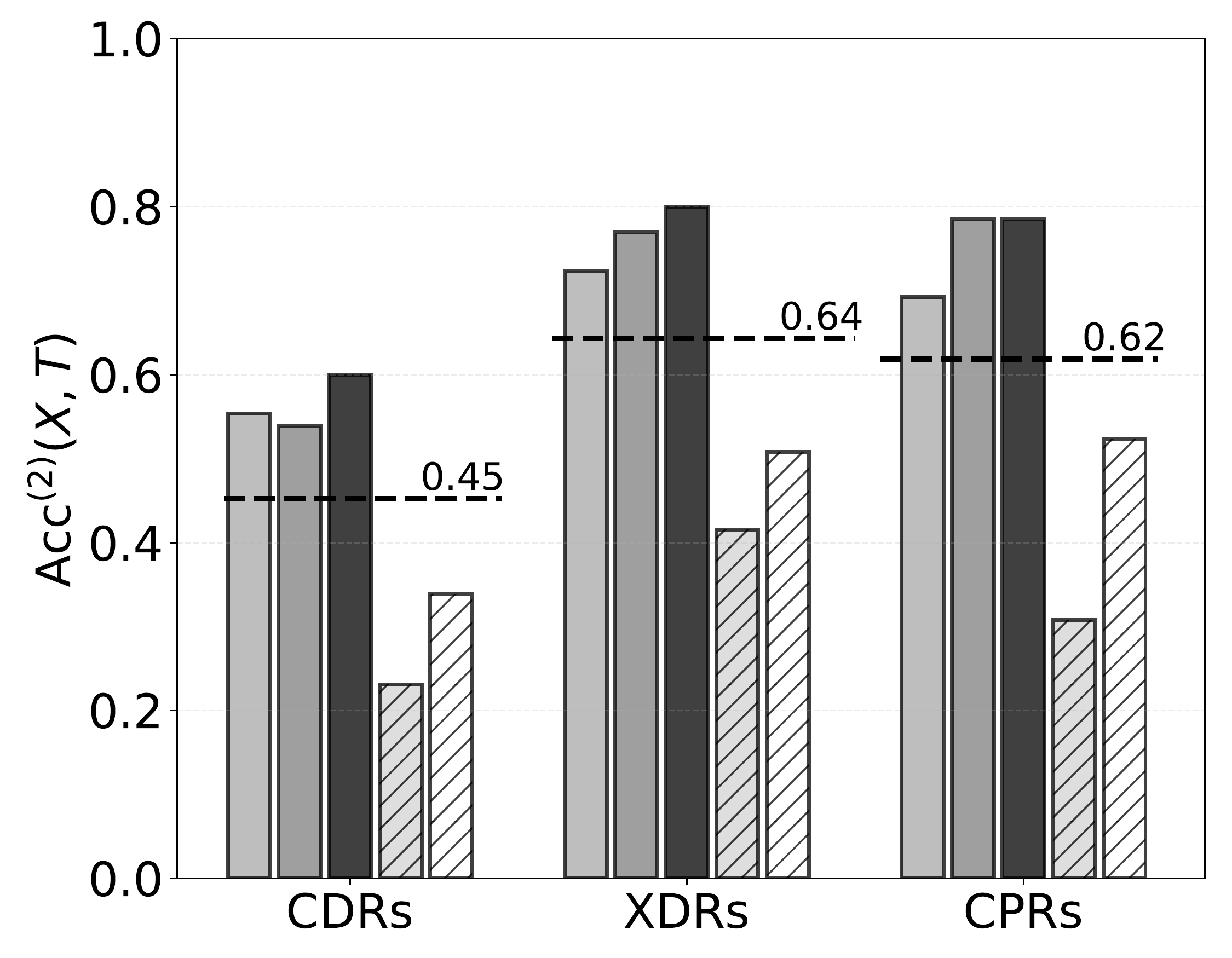}
                \hfill
                \includegraphics[width=0.45\textwidth]{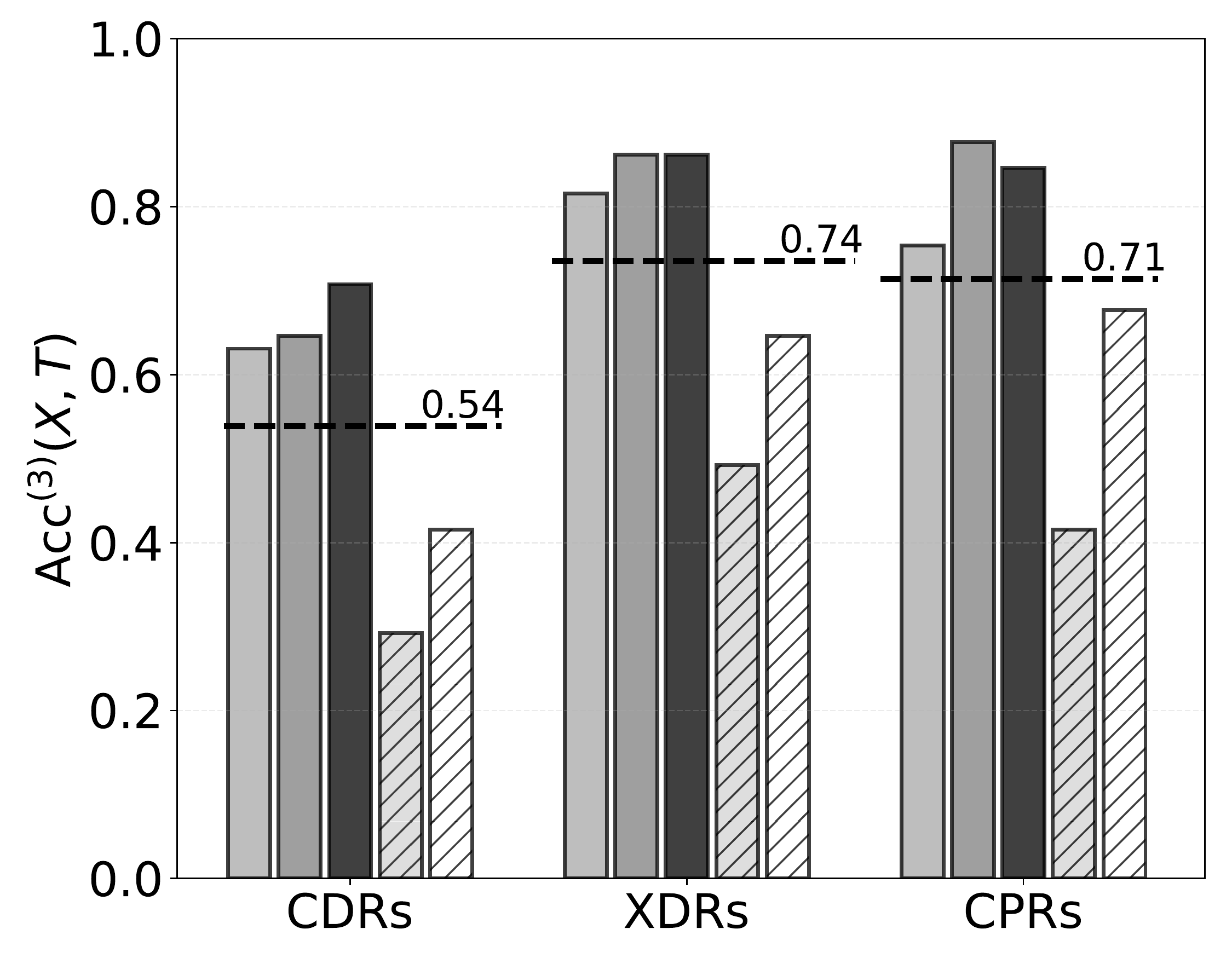}
        \end{subfigure}
    \caption{(Top left) Home Detection Accuracy of the five HDAs, for each data type. (Top right) Accuracy considering just the nearest tower to the user's actual home as ground truth. (Bottom left) 2-accuracy and (Bottom right) 3-accuracy, varying the data type.}
    \label{fig:accuracy_any_tower}
\end{figure}

Figure \ref{fig:accuracy_any_tower} (top left) shows the home detection accuracy of the five HDAs, varying the stream.
We find that, on average, XDRs lead to the highest accuracy ($\overline{\mbox{ACC}}_{X, \mbox{\footnotesize XDRs}} = 0.52$).
HDA1, HDA2, and HDA3 are way far the best algorithms, achieving a home detection accuracy greater than or close to 50\% on XDRs and CPRs, i.e., they can identify the correct home location for more than 50\% of the ground truth users.
In particular, HDA3 is the algorithm achieving the best overall accuracy (ACC$_{\mbox{\footnotesize HDA3}, \mbox{\footnotesize XDRs}} = 0.68 = \mbox{ACC}_{\mbox{\footnotesize HDA3}, \mbox{\footnotesize CPRs}}$).
Figure \ref{fig:accuracy_any_tower} (top right) shows the home detection accuracy if just the closest tower to a user's actual home location (i.e., $H^{(u)} = H_1^{(u)}$) is considered.
Again, XDRs lead to the highest average accuracy and HDA1-3 are way far the best algorithms, even though accuracy drops down to $\overline{\mbox{ACC}}_{X, \mbox{\scriptsize XDRs}} = 0.26$.

\begin{table}
\newcolumntype{Y}{>{\centering\arraybackslash}X}
    \begin{tabularx}{\linewidth}{YY>{\columncolor[gray]{0.8}}cY|Y>{\columncolor[gray]{0.8}}cY}
\toprule
   & \multicolumn{3}{c}{\bf three nearest towers} & \multicolumn{3}{c}{\bf nearest tower}\\
   \cmidrule(l){2-7}
 &  \bf CDRs & \bf XDRs & \bf CPRs  &  \bf CDRs & \bf XDRs & \bf CPRs\\
HDA1 & 0.25 & 0.55 & 0.48 & 0.14 & 0.28 & 0.22\\
HDA2 & 0.35 & 0.63 & 0.69 & 0.2 & 0.32 & 0.26\\
\rowcolor{lightgray}HDA3 & 0.43 & 0.68 & 0.68 & 0.26 & 0.34 & 0.34\\
HDA4 & 0.17 & 0.32 & 0.25 & 0.06 & 0.12 & 0.09\\
HDA5 & 0.26 & 0.43 & 0.37 & 0.09 & 0.22 & 0.18\\
\bottomrule
    \end{tabularx}
    \caption{Accuracy of HDAs by stream.}
    \label{tab:accuracies}
\end{table}

\begin{figure}
    \centering
    \includegraphics[width=1.0\textwidth]{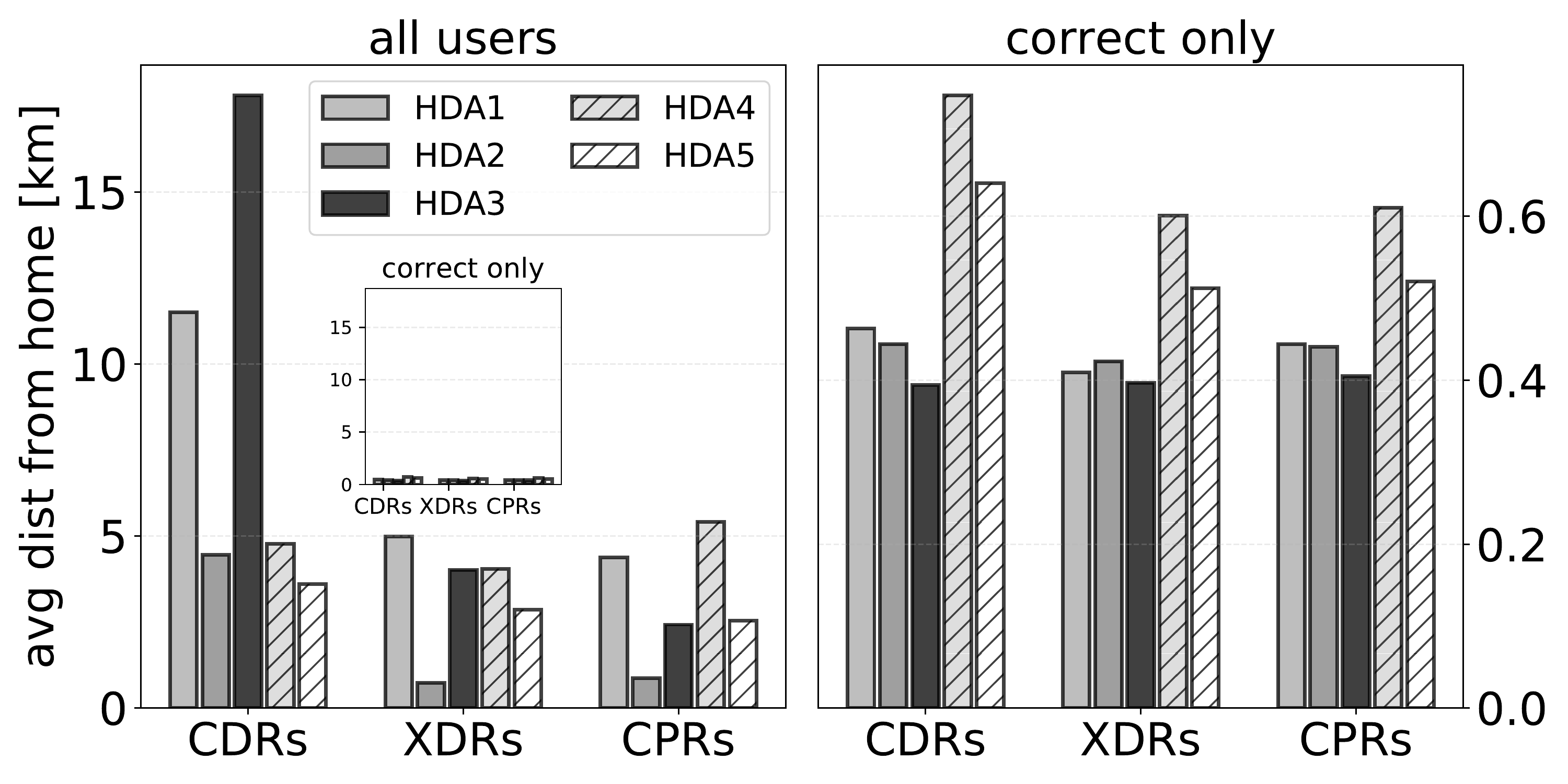}
    \caption{(Left) Average distance between the home location detected by each HDA on each stream and the actual user's home location. (Right) Average distance only for those users for which the HDA correctly detects the home.}
    \label{fig:geoaccuracy}
\end{figure}

The home detection accuracy of an HDA on a stream is computed considering as detected home locations the towers with the highest activity.
We can relax this condition by defining a detected home location to be correct if at least one of the top $k$ towers detected by an HDA $X$ on a stream $T$ is correct.
We denote with $H^{(k)}_{X, T}(u)$ a tower that is within the $k$ towers with the highest activity according to HDA $X$ and stream $T$ for user $u$.
We define the \emph{home detection k-accuracy} (ACC$^{(k)}_{X, T}$) of the combination of HDA $X$ and stream $T$ as the number of $H^{(k)}_{X, T}(u)$ over the total number of ground truth users $|G|$.
Given this definition, ACC$_{X, T} = \mbox{ACC}^{(1)}_{X, T}$.
Figure \ref{fig:accuracy_any_tower} (bottom) shows ACC$^{(k)}_{X, T}$ for $k=2, 3$.
Clearly, ACC$^{(k)}_{X, T}$ increases with $k$ and, again, XDRs lead to the highest average accuracy ($\overline{\mbox{ACC}}^{(2)}_{X, \mbox{\scriptsize XDRs}} = 0.64$, $\overline{\mbox{ACC}}^{(3)}_{X, \mbox{\scriptsize XDRs}} = 0.74$).
Figure \ref{fig:geoaccuracy}(left) shows the average distance of the detected home locations from the individuals' actual home (in km).
Figure \ref{fig:geoaccuracy}(right) does the same but for correctly detected home locations only.
XDRs provides, in general, the best results: on average, the distance between the position of tower detected as the home location has a distance to the actual home lower than 5km.

\subsection{Data Minimization}

We then turn to the question of what is the minimum amount of data, for each stream, that is required to achieve a given sought accuracy, and whether they are qualitatively different at the time of doing so. For instance, does the ``machine-only" feature of CPRs work better than the ``human-machine hybrid" characteristics of XDRs? This experiment is important for both researchers and telco operators: it is usually not easy to access mobile phone data. Thus, requesting these data with a principled approach rather than a "blanket" request, could go a long way in building trust between companies and scientists, while also requiring less human resources and storage infrastructure to prepare and share the data.

\begin{figure}
    \centering
    \begin{subfigure}{\textwidth} 
                \includegraphics[width=0.45\textwidth]{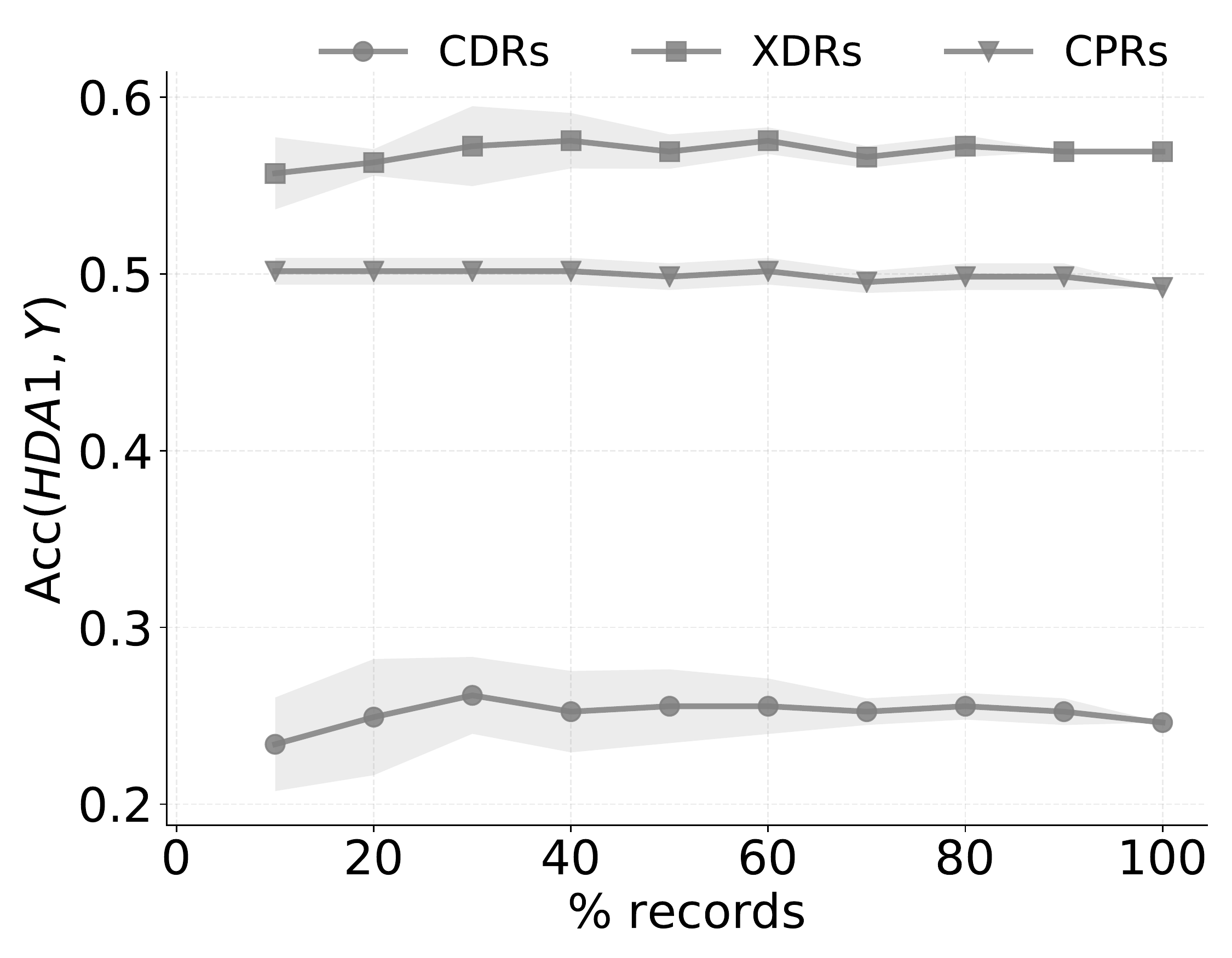}
                \includegraphics[width=0.45\textwidth]{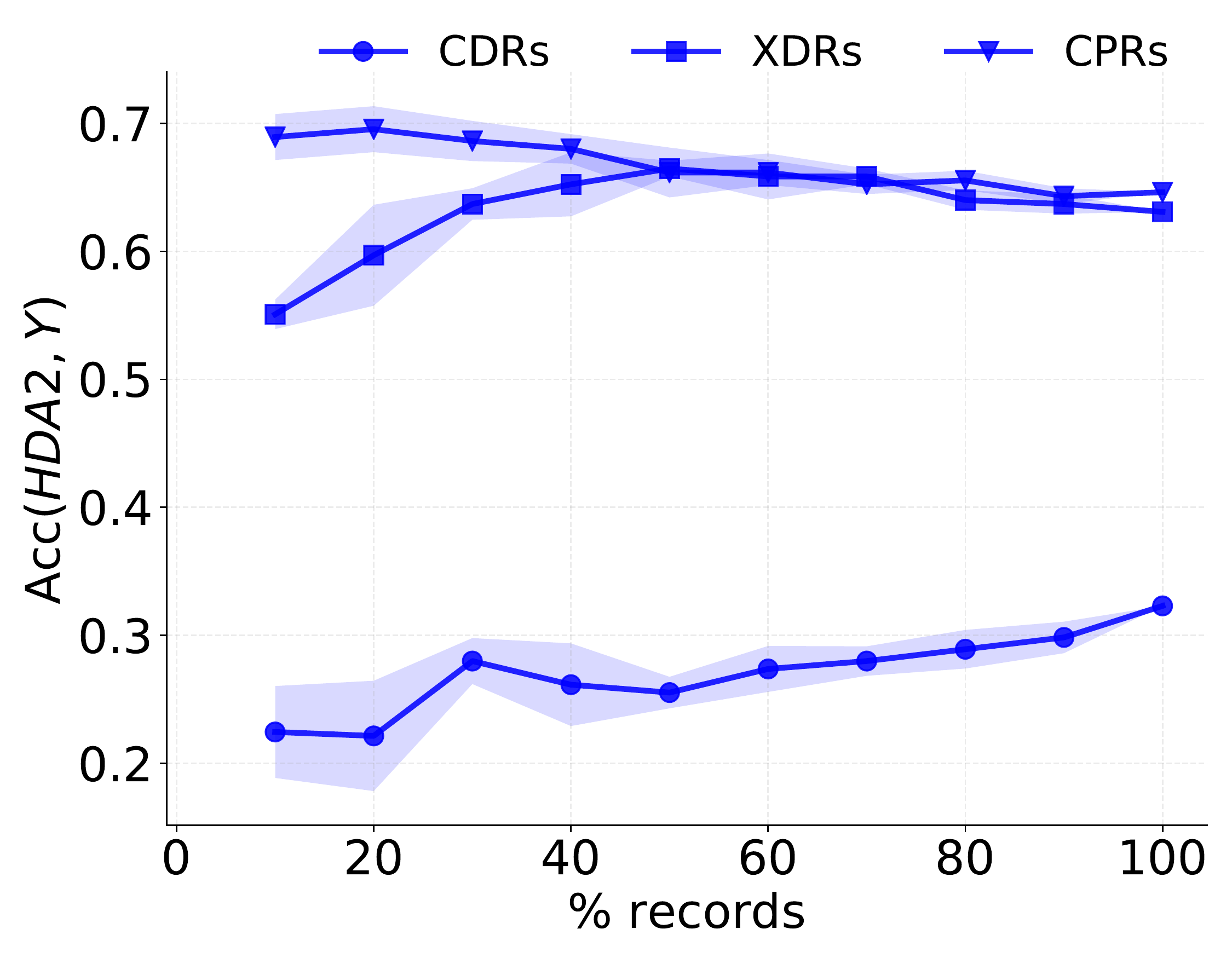}
                \includegraphics[width=0.45\textwidth]{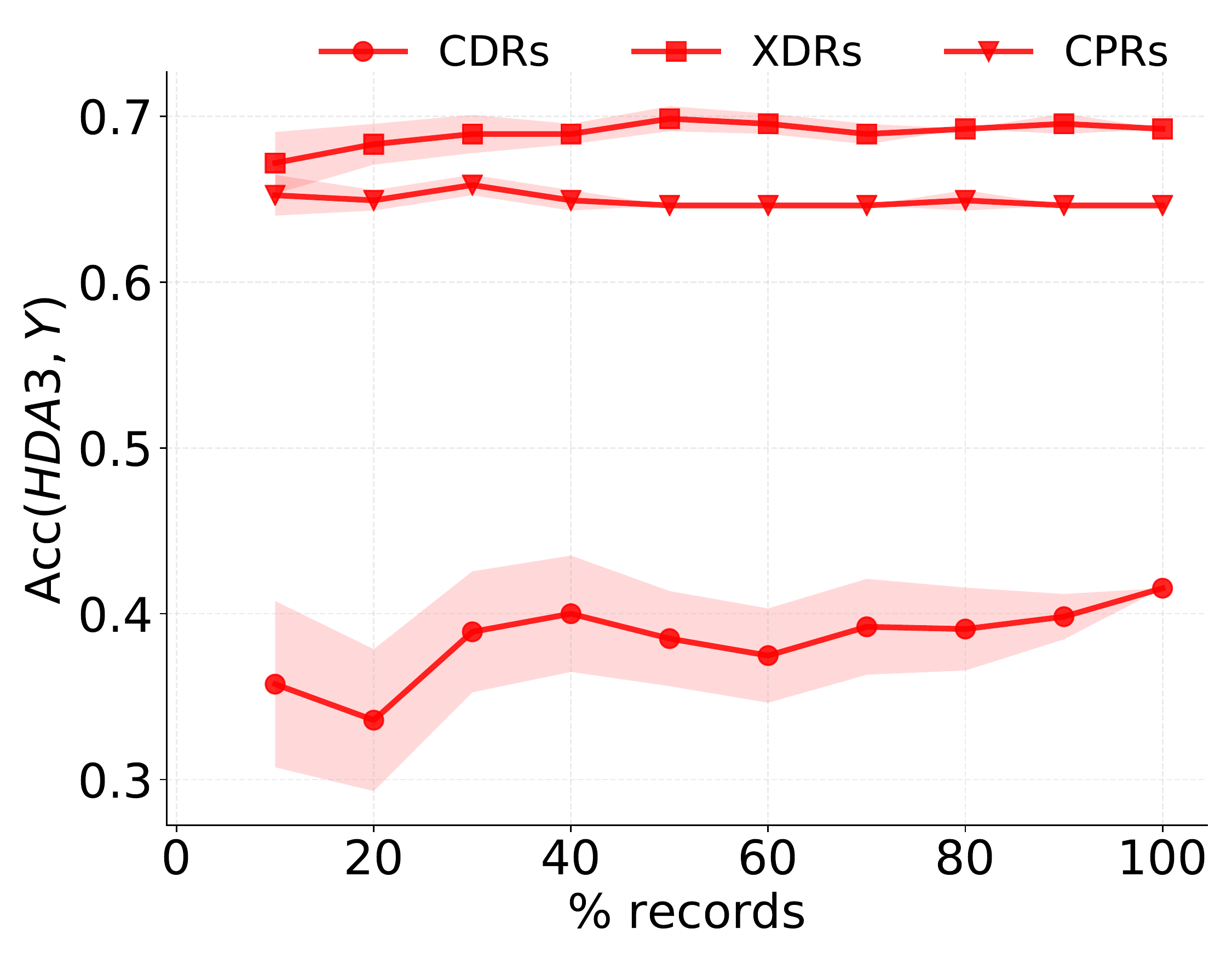}
                \includegraphics[width=0.45\textwidth]{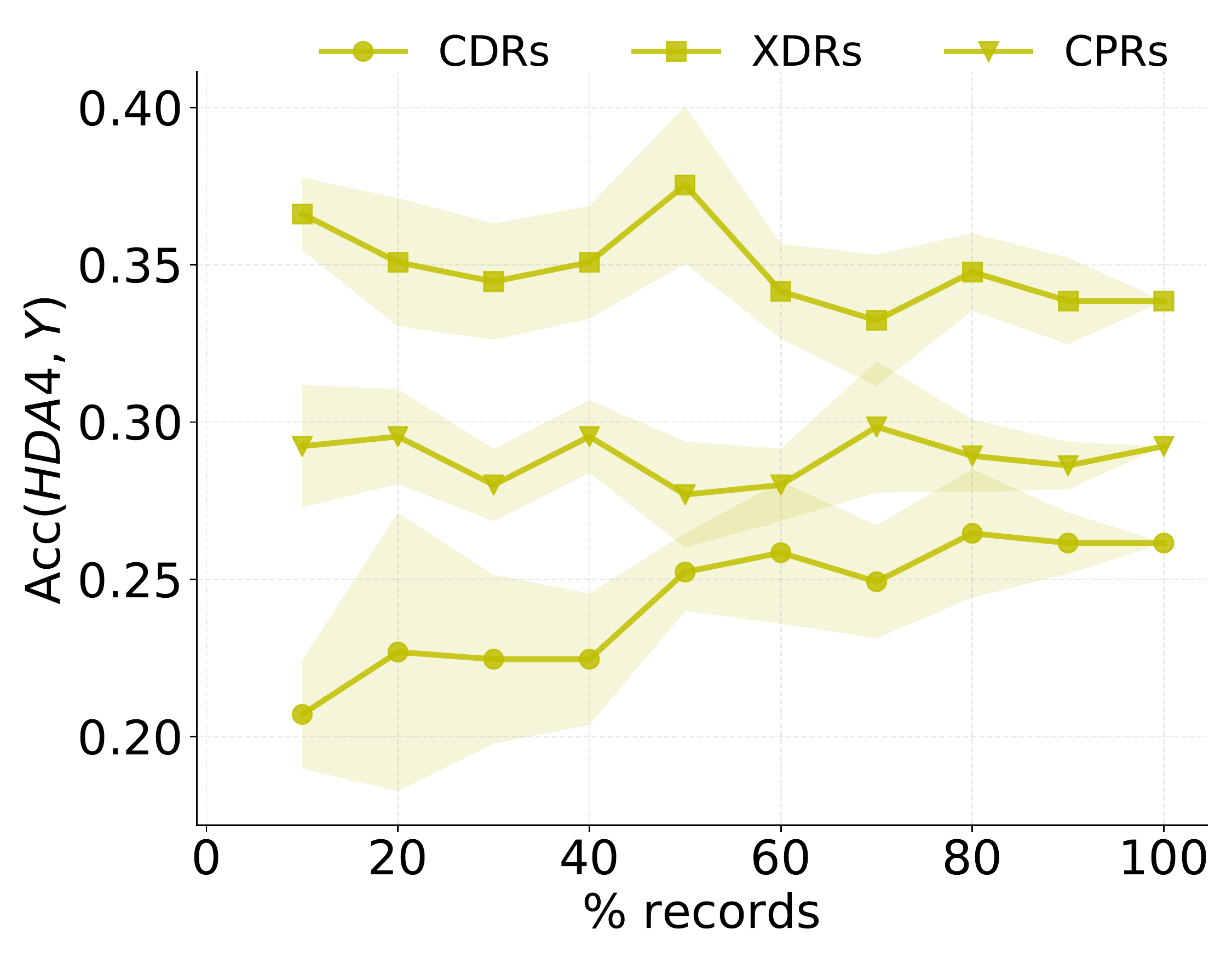}
                \includegraphics[width=0.45\textwidth]{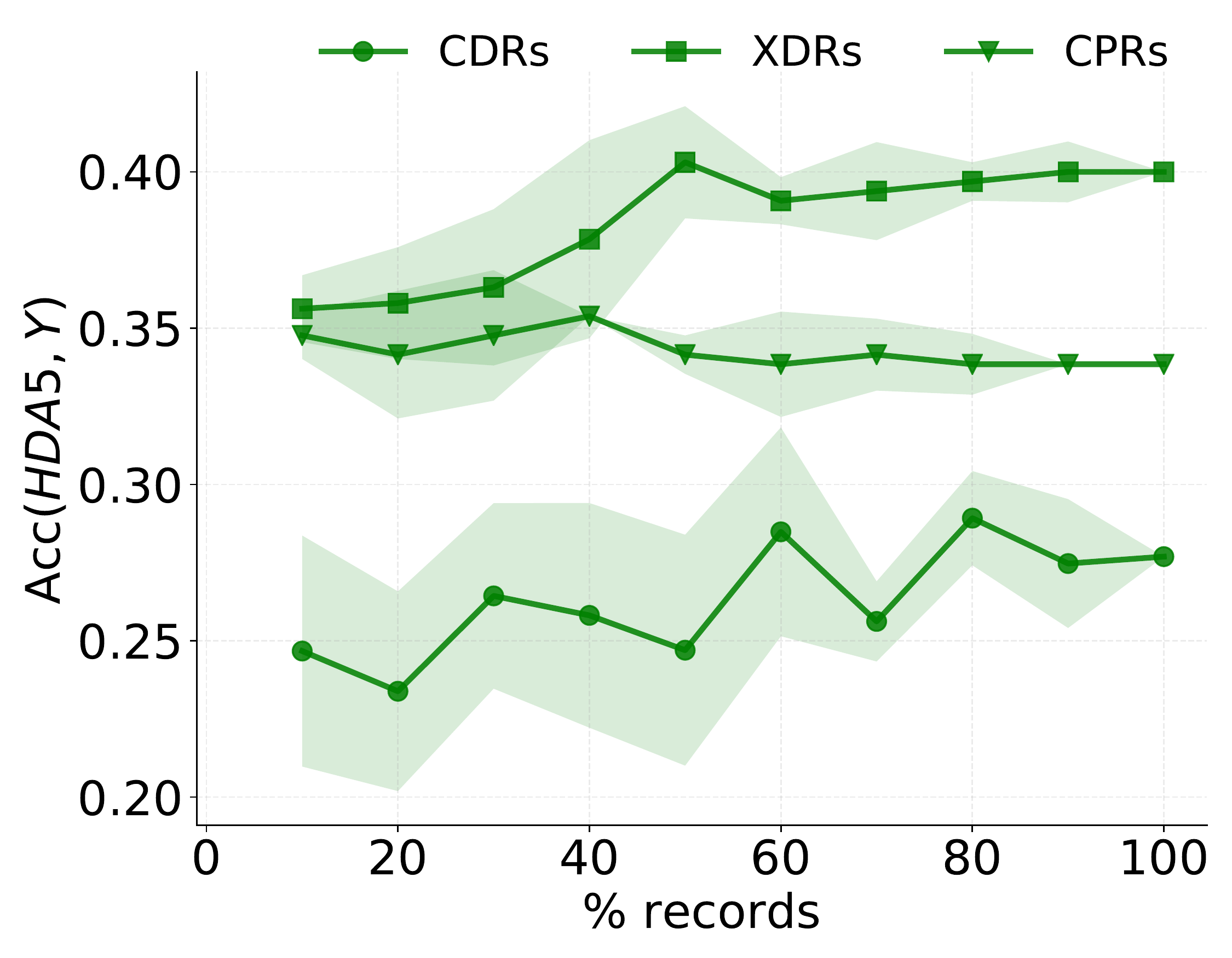}
        \end{subfigure}
    \caption{Results of the data minimization experiment. Number of records randomly selected versus the average and standard deviation of home detection accuracy for HDA1-5.}
    \label{fig:minimization}
\end{figure}

To find out, we run a randomization algorithm that sampled the data at different percentage intervals. Thus, we calculate the home detection algorithms accuracy for 10\%, 20\%, ..., 100\% of their available data without replacement for each user.
We run this experiment five times for each stream to have a more general idea about the data's statistical robustness.
Figure \ref{fig:minimization} shows, for each HDA and stream, the average home detection accuracy varying the number of records used.
The shaded areas are the standard deviation of the different runs (the five runs).

We find three main results. First, CDRs lead to the highest variance in the accuracy across the trials: the accuracy of home detection can largely vary based on the set of records selected per user. In contrast, CPRs lead to the most stable accuracy.  Second, a high fraction of CDRs is needed to achieve an accuracy similar to that obtained using 100\% of the records. For example, for HDA3, at least 40\% of the records are needed. In contrast, just a small fraction of CPRs (even 10\%) is enough to achieve the same accuracy as using all the records. XDRs lie between these two extremes.

\section{Conclusions}
\label{sec:summary-conclusions}
Mobile phone data are a crucial data source for official statistics, including the important task of home location detection.
To the best of our knowledge, this is the first time that individual-level ground truth data are used to assess the accuracy of the existing HDAs.
Our analysis reveals that the type of stream used - CDRs, XDRs, CPRs - heavily influences the home detection accuracy.
Similarly, the choice of the algorithm is crucial: HDA3, which defines the home of an individual as the most active tower during nighttime, is the most accurate algorithm, regardless of the stream used.
Consequently, using XDRs in combination with HDA3 is the solution leading to the highest home detection accuracy.
Nevertheless, our data minimization experiment revealed that CPRs are the most resilient to reducing the number of records available per user, and XDRs the second most resilient stream.
Overall, our work demonstrates that CDRs, the most used stream in the literature to detect individuals' homes, lead to low accuracy and stability, questioning the plethora of research made in the last years.

Experiences like ours may contribute to shaping the discussion on the best mobile phone stream to capture presences and human mobility patterns.
This is crucial because citizens and policy makers' decisions depend on what we measure, how right our measurements are, and how well our measures are understood.

\section*{Acknowledgements}
Luca Pappalardo has been partially funded by EU project SoBigData++
RI, grant \#871042.  L.F. thanks V\'ictor Navarro, and acknowledges the funding and support of Telef\'onica R\&D Chile and CISCO Chile.
Ciro Cattuto acknowledges partial support from the Lagrange Project of ISI Foundation funded by CRT Foundation.

\section*{Author contributions}
LP analyzed the data, implemented and ran the home detection algorithms, set up the experiments, and made the plots.
\\LF analyzed the data, implemented and ran the home detection algorithms, set up the experiments, and made the plots.
\\MS mined and provided the anonymized mobile phone data.
\\CC analyed the data and set up the experiments.
\\LB mined and provided the anonymized mobile phone data.
\\All authors interpreted the results, wrote and approved the manuscript.

\section*{Competing interests}
The authors declare no competing interests.

\bibliographystyle{plain}
\bibliography{bib.bib}

\end{document}